\documentclass[%
 reprint,
nofootinbib, 
 amsmath,amssymb,
 aps,
nofootinbib
]{revtex4-2}
\skip\footins 0.5cm

\usepackage{graphicx}
\usepackage{dcolumn}
\usepackage{bm}
\usepackage{mathcomp}
\RequirePackage{textpos}

\usepackage{caption} 
\usepackage{subcaption}

\begin{document}

\preprint{APS/123-QED}

\title{ Measurement of the H to ZZ branching fraction at 350 GeV and 3 TeV CLIC }

\author{N. Vuka\u{s}inovi\'{c}}
 \email{nvukasinovic@vin.bg.ac.rs}

\author{I. Bo\v{z}ovi\'{c}-Jelisav\u{c}i\'{c}}%
\author{G. Ka\u{c}arevi\'{c}}%
\author{G. Milutinovi\'{c}-Dumbelovi\'{c}}%
\author{T. Agatonovi\'{c}-Jovin}%
\author{I. Smiljani\'{c}}%
\affiliation{``VIN\u{C}A'' Institute of Nuclear Sciences - National Institute of the Republic of Serbia, University of Belgrade, Belgrade, Serbia
}
\collaboration{CLICdp Collaboration}

\author{M. Radulovi\'{c}}
\author{J. Stevanovi\'{c}}

\affiliation{University of Kragujevac, Faculty of Science, Kragujevac, Serbia
}

\date{\today}

\begin{abstract}
In this paper we investigate the prospects for measuring the branching fraction of the Standard Model Higgs boson decay into a pair of $Z$ bosons at the future Compact Linear Collider (CLIC) at 350\,GeV and 3\,TeV centre-of-mass energies. Studies are performed using a detailed simulation of the detector for CLIC, taking into consideration all relevant physics and beam-induced background processes. It is shown that the product of the Higgs production cross-section and the branching fraction BR(${H\rightarrow\thinspace ZZ^\ast}$) can be measured with a relative statistical uncertainty of 20\% (3.0\%) at a centre-of-mass energy of 350\,GeV (3\,TeV) using semileptonic final states, assuming an integrated luminosity of 1\,ab$^{-1}$ (5\,ab$^{-1}$). 
\end{abstract}

\maketitle


\section{Introduction}
As a staged $e^+e^-$ collider, CLIC can provide a comprehensive physics programme of measurements in the Higgs sector. The large samples of data accumulated from all energy stages enables precise measurements of the Higgs couplings, mass and width. Centre-of-mass energies above 1\,TeV enhance measurements of the Higgs self-coupling as well as the sensitivity to probe Beyond the Standard Model physics (BSM) in the Higgs sector. As it is designed to operate at the highest centre-of-mass energies of any proposed $e^+e^-$ collider project, and having the option of up to 80\% electron beam polarisation, CLIC offers an extensive set of key physics measurements of the Higgs sector.

In general, it is important to measure the Higgs couplings with the highest possible precision. Most of the BSM models predict Higgs couplings to electroweak bosons to deviate from the Standard Model (SM) predictions at the order of a percent \cite{r1}. As discussed in \cite{r2}, a global fit to data from all energy stages allows extraction of the Higgs couplings with the required precision.

So far, BR(${H\rightarrow\thinspace ZZ^\ast}$) has only been studied in detail at 1.4\,TeV centre-of-mass energy \cite{goca}, with an estimate made for 3\,TeV centre-of-mass energy on the basis of luminosity scaling. Although of lower precision than the high-energy measurements, the 350\,GeV data will complete the set of Higgs branching fraction measurements at CLIC, serving as input to a global fit of the Higgs couplings in the Effective Field Theory (EFT) approach \cite{r3}.

In this paper we determine the CLIC statistical precision to measure $H\rightarrow ZZ^*$ branching ratio at 350\,GeV and 3\,TeV centre-of-mass energies in the semileptonic final state, using the full simulation of experimental conditions. The semileptonic final state is chosen because its irreducible background is lower than that of the hadronic final state.
 
The paper is organized as follows: A detector for CLIC is described in Section 2; Section 3 lists possible Higgs production mechanisms at CLIC, while Sections 4 to 6 provide details on event samples, the analysis methods and predicted statistical precision of the measurements.

\section{\label{sec:leve2}The CLIC\_ILD detector model}
The CLIC\_ILD detector \cite{r4}, based on the ILD detector concept for ILC \cite{r5}, has been modified for the experimental conditions at CLIC. More recently, the CLICdet detector concept \cite{r6} has been developed. Both detector concepts use fine-grained electromagnetic and hadronic calorimeters \break(ECAL and HCAL) optimized for the Particle Flow Algorithm (PFA) employed in event reconstruction \cite{r7}. Muon momentum resolution is required to be $\sigma{ (p_{t}/p^{2}_{t})}$ $\sim$ 2 $\cdot$ 10$^{-5}$ GeV$^{-1}$ \cite{r4}, while the jet-energy resolution ranges between $3.5\%$ and $5\%$ depending on the jet energy \cite{r4}. The later is considered crucial for separation of nearby jets from Higgs, $W$ and $Z$ bosons. Differences between the detector models are found to have no significant impact on the statistical precision of the measurements discussed in this paper.

\section{\label{sec:leve3} Higgs production mechanisms at CLIC }

\begin{figure*}
\centering
\includegraphics[width=0.8\columnwidth]{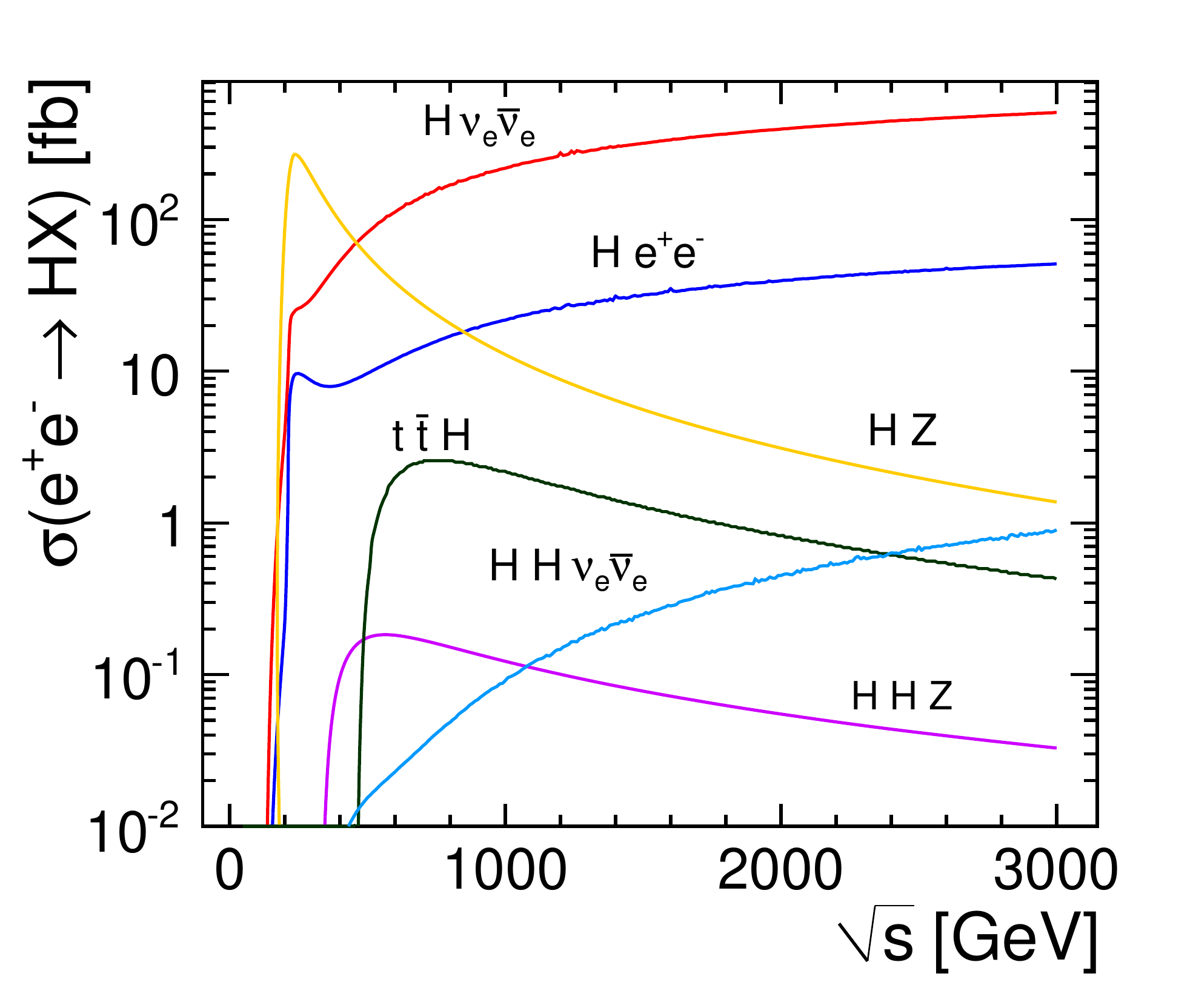}
\caption{\label{fig-ww} Unpolarized cross-sections as a function of centre-of-mass energy for the main Higgs production processes at an $e^{+}e^{-}$ collider, assuming a Higgs boson mass of 126 GeV \cite{r2}.}
\end{figure*}

CLIC operation is expected to be staged at three centre-of-mass energies: 380 (350)\,GeV, 1.5\,TeV and 3\,TeV. \hspace{7cm} The currently anticipated lowest energy stage of CLIC is 380 GeV\footnote{380\,GeV centre-of-mass energy is considered the optimum energy for the first stage as it enables both Higgsstrahlung and top-quark measurements above the t$\bar{\mathrm{t}}$ threshold.}. The studies presented in this paper are performed at 350 GeV, with results scaled to the updated integrated luminosities at 380\,GeV from \cite{r8}. The first stage enables precision measurement in both the Higgs and top-quark sectors. Taking into account beam polarisation, CLIC will produce about 4.5\,$\cdot$\,10$^{6}$ Higgs bosons combining data from all energy stages \cite{r9}. As illustrated at Figure \ref{fig-ww} \cite{r2}, the main Higgs production mechanism in the first stage is Higgsstrahlung ($HZ$), while at around 500 GeV centre-of-mass energy $WW$-fusion ($H\nu_{e}\bar{\nu_{e}}$) starts to dominate. The cross-section for the Higgsstrahlung process at 350 GeV is 129 fb, while at 3 TeV the cross-section for Higgs production in $WW$-fusion is 415.05 fb. The branching fraction for the $H\rightarrow ZZ^*$ decay is 2.89\% \cite{r10}.  
The expected number of $HZ$ events in which the primary $Z$ decays hadronically is around 9.3 $\cdot$ $10^{4}$ in 1\,ab$^{-1}$ of unpolarized data. The expected number of H$\nu_{e}\bar{\nu_{e}}$ events is around 2 $\cdot$ $10^{6}$ in 5\,ab$^{-1}$ of unpolarized data. The above estimates assume a realistic CLIC luminosity spectrum with Initial State Radiation (ISR) included.
The CLIC accelerator baseline design foresees sharing the running time for $-$80 \% and $+$80\% $e^{-}$ polarization in the ratio 80:20 at 1.5\,TeV and 3\,TeV, while the ratio 50:50 is assumed at 380\,GeV, with no $e^{+}$ polarization at any stage \cite{r8}. These assumed polarization schemes will collectively be referred to as \textit{beam polarisation} throughout the text. Due to the chiral nature of the charged-current interaction, $WW$-fusion is much more affected by the $e^{-}$ polarization than is the Higgsstrahlung process. With the proposed polarization scheme, the cross-section for $WW$-fusion will increase by a factor of $\sim$ 1.5 \cite{r8}. The impact of the beam polarisation on the statistical precision of the ${\sigma(H\nu\bar{\nu})\times BR(H\rightarrow ZZ^\ast)}$ measurement at 3\,TeV centre-of-mass energy is discussed in Section 6.



\section{\label{sec:leve4} Event samples and preselection}

\subsection{Event samples}

Signal and background events are simulated using the \break Whizard 1.95 event generator \cite{r11}. The processes of\break hadronization and fragmentation of final-state quarks and gluons are simulated using Pythia 6.4 \cite{r12}. The Higgs boson mass is assumed to be 126 GeV in the simulations. The CLIC luminosity spectrum and interactions between beams are obtained using GuineaPig 1.4.4 \cite{r13}, while hadron production from Beamstrahlung photons is simulated with Pythia 6.4. List of the signal and background processes considered are given in Table \ref{table:1a} and Table \ref{table:1b}, at 350 GeV and 3 TeV, respectively. Note that processes involving photons from Beamstrahlung are not considered as a background at 350 GeV due to the fact that these processes are much less pronounced at lower centre-of-mass energies and thus contribute negligibly to this study. However, in order to simulate a realistic experimental environment at CLIC, the hadronic background from Beamstrahlung is overlaid before the digitisation phase on the reconstructed signal and background events at all centre-of-mass energies. At 3 TeV, simulation of the background process $e^-e^+\rightarrow q\bar{q}l^+l^-\nu\bar{\nu}$ was available only at the generator level. Approximately\break 99.8\% of these events can be removed by considering optimised intervals of the Higgs mass and off-shell $Z$ mass. It is estimated that fewer than 30  $q\bar{q}l^+l^-\nu\bar{\nu}$ events will remain in 5\,ab$^{-1}$ of data, which has a negligible impact on the statistical uncertainty of the branching fraction measurement.     

Interactions with the detector are simulated using the CLIC\_ILD detector model within the Mokka simulation package \cite{r14} using the GEANT4 framework \cite{r15}. Event reconstruction is based on the Particle Flow Algorithm (PFA) implemented in the Pandora toolkit \cite{r16}. Particles are reconstructed as particle-flow objects (PFOs) by combining the information from different sub-detectors. For the jet clustering, the $k_{T}$ algorithm \cite{r17} is used in the exclusive mode, implemented in the FastJet processor \cite{r18}. The Isolated Lepton Finder Marlin processor \cite{r19} is used for isolated lepton ($e, \mu$) identification. Tagging of beauty and charm jets is performed with the LCFIPlus processor \cite{r20}. The TMVA package \cite{r21} is used for the multivariate classification (MVA) of signal and background events using their kinematic properties. The simulation, reconstruction and analyses are carried out with the ILCDIRAC framework \cite{r22}.

\begin{table}
\centering
\caption[Caption for LOF] %
{Processes considered with the corresponding cross-sections, expected number of events and simulated sample size ($N_{\mathrm{sim}}$) at 350 GeV (a) and 3 TeV (b) centre-of-mass energies. For signal candidates, $N_{\mathrm{sim}}$ only includes events having two truth-linked\footnotemark[1] leptons (electrons or muons).} \label{table:1} 
\begin{subtable} {\columnwidth} 
\caption{} \label{table:1a} 
\begin{tabular*}{1.\columnwidth}{@{\extracolsep{\fill}}llllll@{}}
\hline 
\multicolumn{1}{@{}l}{Signal process}    & {$\sigma$(fb)}   & {$N @ 1 \mathrm{ab^{-1}}$}  & {$N_{\mathrm{sim}}$}  \\
\hline
\begin{tabular}{p{3.5cm}p{3cm}}$e^-e^+\rightarrow HZ; Z\rightarrow q\bar{q}, H\rightarrow ZZ^\ast, ZZ^\ast\rightarrow q\bar{q}l^+l^- (l = e, \mu)$\end{tabular}              & 0.24   & 240    & 17721  \\
\hline 
\multicolumn{1}{@{}l}{Background processes}    & \multicolumn{1}{p{0.03cm}}{$\sigma$(fb)}   & \multicolumn{1}{p{1.9cm}}{$N @ 1 \mathrm{ab^{-1}}(\cdot 10^{3})$}  & \multicolumn{1}{p{1.3cm}}{ $N_{\mathrm{sim}}(\cdot 10^{3})$}  \\
\hline
\begin{tabular}{p{3cm}p{3cm}}$e^-e^+\rightarrow HZ; Z\rightarrow q\bar{q}, H\rightarrow others$\end{tabular}                     &  7.0   & 7   & 77   \\
\begin{tabular}{p{3cm}p{3cm}}$e^-e^+\rightarrow HZ; Z\rightarrow q\bar{q}, H\rightarrow WW\rightarrow 4q$\end{tabular}       & 10.5  & 10.5     & 12  \\
\begin{tabular}{p{3cm}p{3cm}}$e^-e^+\rightarrow HZ; Z\rightarrow \mu^{+}\mu^{-}, H\rightarrow others$\end{tabular}                      & 2.3   & 2.3        & 85    \\
\begin{tabular}{p{3cm}p{3cm}}$e^-e^+\rightarrow HZ; Z\rightarrow e^{+}e^{-}, H\rightarrow others$\end{tabular}                      & 2.3  & 2.3 	    & 85  \\
\begin{tabular}{p{3cm}p{3cm}}$e^-e^+\rightarrow HZ; Z\rightarrow \mu^{+}\mu^{-}, H\rightarrow WW\rightarrow 4q$\end{tabular}                        & 0.7  & 0.7     & 14 \\
\begin{tabular}{p{3cm}p{3cm}}$e^-e^+\rightarrow HZ; Z\rightarrow e^{+}e^{-}, H\rightarrow WW\rightarrow 4q$\end{tabular}                      & 0.7 & 0.7  & 14   \\
\begin{tabular}{p{3cm}p{3cm}}$e^-e^+\rightarrow q\bar{q}q\bar{q}l^+l^-$\end{tabular}                        & 4.5 & 4.5  & 44    \\
\begin{tabular}{p{3cm}p{3cm}}$e^-e^+\rightarrow qqqq$\end{tabular}                 & 5847 & 5.8 $\cdot$ $10^{3}$ & 191     \\
\begin{tabular}{p{3cm}p{3cm}}$e^-e^+\rightarrow q\bar{q}l^+l^-$\end{tabular}                   & 1704 & 1.7 $\cdot$ $10^{3}$ & 746   \\
\hline
\hline
\end{tabular*}
\end{subtable}
\quad
\begin{subtable} {\columnwidth} 
\vspace*{0.5 cm}
\centering
\caption{} \label{table:1b} 
\begin{tabular*}{1.\columnwidth}{@{\extracolsep{\fill}}llllll@{}}
\hline 
\multicolumn{1}{@{}l}{Signal process}    & {$\sigma$(fb)}   & {$N @ 5 \mathrm{ab^{-1}}$}  & {$N_{\mathrm{sim}}$}  \\
\hline
\begin{tabular}{p{3cm}p{3cm}}$e^-e^+\rightarrow H\nu\bar{\nu}; H\rightarrow ZZ^\ast, ZZ^\ast\rightarrow q\bar{q}l^+l^-,\break (l = e, \mu)$\end{tabular}              & 1.13   & 5650    & 16752  \\
\hline
\multicolumn{1}{@{}l}{Background processes}    & \multicolumn{1}{p{0.05cm}}{$\sigma$(fb)}   & \multicolumn{1}{p{1.8cm}}{$N @ 5 \mathrm{ab^{-1}}(\cdot 10^{3})$}  & \multicolumn{1}{p{0.9cm}}{ $N_{\mathrm{sim}}(\cdot 10^{3})$}  \\
\hline 
\begin{tabular}{p{3cm}p{3cm}}$e^-e^+\rightarrow H\nu\bar{\nu}; H\rightarrow WW, WW\rightarrow 4q$\end{tabular}       & 43  & 218     & 219  \\
\begin{tabular}{p{3cm}p{3cm}}$e^-e^+\rightarrow H\nu\bar{\nu}; H\rightarrow b\bar{b}$\end{tabular}                      & 233   & 1.2 $\cdot$ $10^{3}$       & 1.1 $\cdot$ $10^{3}$   \\
\begin{tabular}{p{3cm}p{3cm}}$e^-e^+\rightarrow H\nu\bar{\nu}; H\rightarrow c\bar{c}$\end{tabular}                      & 11.7  & 58.5     &52  \\
\begin{tabular}{p{3cm}p{3cm}}$e^-e^+\rightarrow H\nu\bar{\nu}; H\rightarrow gg$\end{tabular}                        & 35.2  & 176     & 128 \\
\begin{tabular}{p{3cm}p{3cm}}$e^-e^+\rightarrow H\nu\bar{\nu}; H\rightarrow others$\end{tabular}                   & 91  & 452    & 465 \\
\begin{tabular}{p{3cm}p{3cm}}$e^-e^+\rightarrow q\bar{q}l^+l^-$\end{tabular}                      & 3320 & 16.6 $\cdot$ $10^{3}$ & 2 $\cdot$ $10^{3}$  \\
\begin{tabular}{p{3cm}p{3cm}}$e^-e^+\rightarrow qql\nu$\end{tabular}                        & 5561 & 27.8 $\cdot$ $10^{3}$ & 3.1$\cdot$ $10^{3}$   \\
\begin{tabular}{p{3cm}p{3cm}}$e^-e^+\rightarrow q\bar{q}\nu\bar{\nu}$\end{tabular}                     &  1317   & 6.6 $\cdot$ $10^{3}$  & 569   \\
\begin{tabular}{p{3cm}p{3cm}}$\gamma\gamma\rightarrow q\bar{q}l^+l^-$\end{tabular}                 & 20293 & 135.7 $\cdot$ $10^{3}$ & 2.5 $\cdot$ $10^{3}$    \\
\begin{tabular}{p{3cm}p{3cm}}$\gamma\gamma\rightarrow q\bar{q}$\end{tabular}                   & 112039 & 517.4 $\cdot$ $10^{3}$ & 1 $\cdot$ $10^{3}$  \\
\begin{tabular}{p{3cm}p{3cm}}$e^\pm\gamma\rightarrow q\bar{q}e$\end{tabular}   & 20661 & 60.3 $\cdot$ $10^{3}$ & 462     \\
\begin{tabular}{p{3cm}p{3cm}}$e^\pm\gamma\rightarrow qq\nu$\end{tabular}    & 36832 & 138.3 $\cdot$ $10^{3}$ & 692     \\
\begin{tabular}{p{3cm}p{3cm}}$e^-e^+\rightarrow q\bar{q}l^+l^-\nu\bar{\nu}$\end{tabular}                     &  3.4   & 17   & 10   \\
\hline 
\hline
\end{tabular*}
\footnotetext[1]{Truth-linking refers to association of the reconstructed particles, in this case of a lepton pair, with generated decay products of a Higgs boson.\\}
\end{subtable}
\end{table}

\subsection{Preselection}

The analyses consist of a loose preselection followed by a MVA based selection. The preselection requirement for measurements at both 350 GeV and 3 TeV is that exactly two isolated leptons of the same flavour and opposite charge (electrons or muons) are found per event. Lepton isolation is optimized according to track energy, the ratio of energies deposited in the electromagnetic and hadronic calorimeters as well as the impact parameters of the lepton tracks.  

\begin{figure}[h]
\centering
\includegraphics[width=\columnwidth]{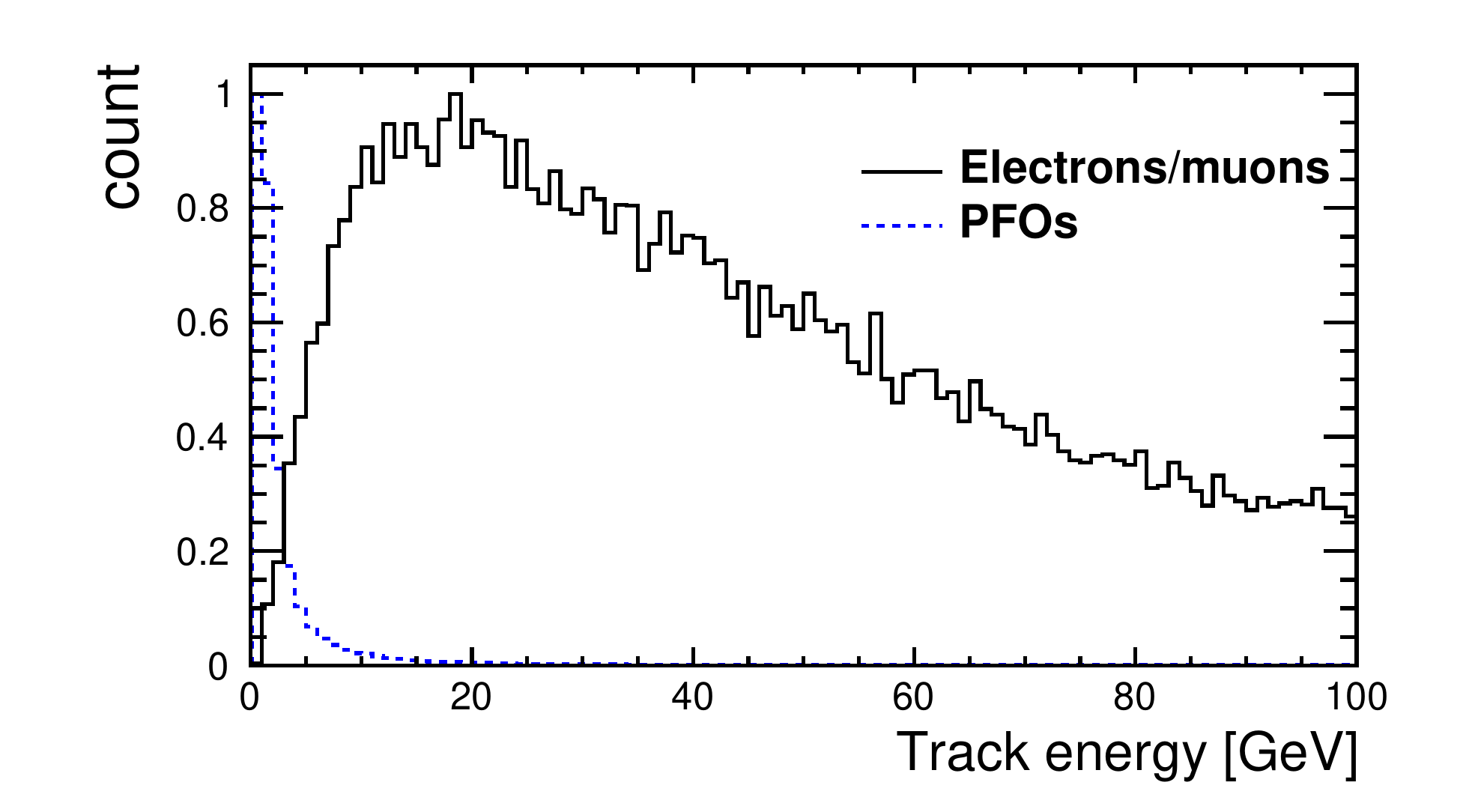}
\begin{textblock}{0.08}(4.7,-2.4) 
\textbf{CLICdp}
\end{textblock}
\caption{\label{fig-trackenergy}Energy of the reconstructed signal leptons (solid) and other reconstructed PFO objects (dashed) in signal events, at 3 TeV centre-of-mass energy. }
\end{figure}

Electrons and muons originating from $ZZ^{*}$ decays have energies that are much higher than the energy of a typical PFO in a jet, as illustrated in Figure \ref{fig-trackenergy}. The selection is optimized in such a way that muons and electrons are required to have an energy of at least 5\,GeV (6\,GeV) at 350\,GeV (3\,TeV).

Charged leptons from the Z decay are required to be consistent with production at the primary vertex. Due to this fact the range of impact parameter components has been optimized as well. The 3-d ($R_{0}$) impact parameter can be decomposed into longitudinal ($z_{0}$) and transverse ($d_{0}$) components. In a signal event, electrons and muons will have significantly smaller impact parameters than other reconstructed particles\footnote{This is particularly the case if compared with impact parameter of Beamstrahlung products ($\gamma_{\mathrm{BS}}\gamma_{\mathrm{BS}}\rightarrow hadrons$) or particles from heavy quark jets from $Z\rightarrow q\bar{q}$ decays.}. Thus it is required: $d_{0}<$ 0.02\,mm and $z_{0}<$ 0.02\,mm at 350\,GeV and $d_{0}<$ 0.02\,mm, $z_{0}<$ 0.03\,mm and $R_{0}<$ 0.03\,mm at 3\,TeV centre-of-mass energies. 


Muons can be distinguished from electrons using the ratio $R_{CAL}$ of energy deposits in ECAL and HCAL:
\begin{equation}
\label{eq-RCAL}
R_{CAL} =E_{ECAL}/(E_{ECAL}+E_{HCAL})
\end{equation}

\begin{figure}
\centering
\includegraphics[width=\columnwidth]{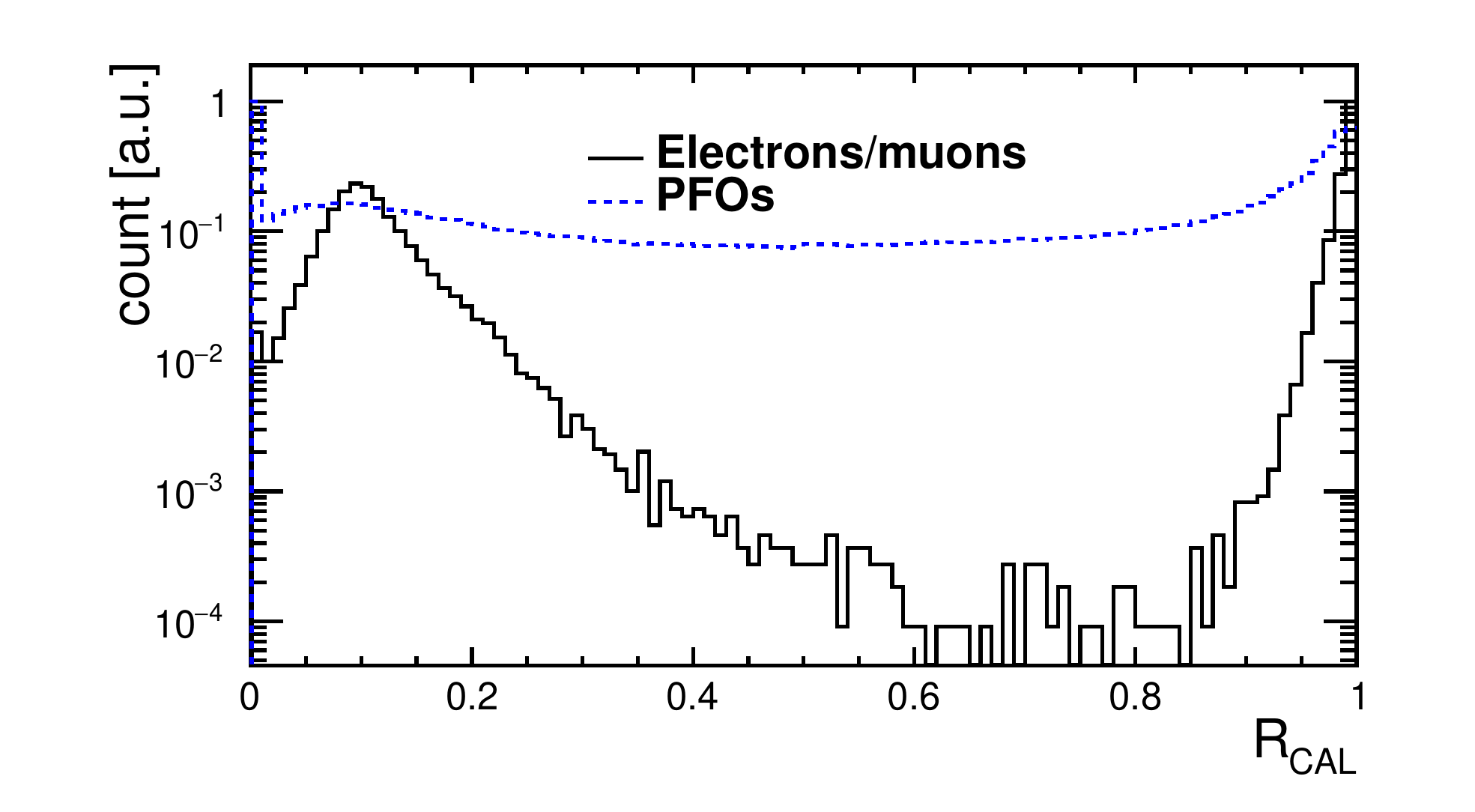}
\begin{textblock}{0.08}(4.7,-2.3)
\textbf{CLICdp} 
\end{textblock}
\caption{\label{fig-rcal}Calorimeter energy ratio $R_{CAL}$  (Eq.\ref{eq-RCAL}) of reconstructed electrons and
muons (solid) and other reconstructed particles (dashed), for signal at 3 TeV centre-of-mass energy.}
\end{figure}

\noindent Because electrons are contained within the ECAL, they peak at $R_{CAL}$ = 1. Muons deposit a minimal amount of energy throughout the calorimeters and have a peak at $R_{CAL}$ = 0.1. This is illustrated in Figure \ref{fig-rcal}, for reconstructed signal at 3 TeV centre-of-mass energy. In order to remove particles which do not behave as electrons or muons in the calorimeters, the calorimeter energy ratio $R_{CAL}$ is required to be: (0.35 $<R_{CAL}< $ 0.9) at 350 GeV and ($R_{CAL} > $ 0.94) or (0.02 $ < R_{CAL} < $ 0.35) at 3\,TeV centre-of-mass energies. 

\begin{figure}  
\begin{subfigure} {\columnwidth}
\caption{}\label{fig:4a}
\centering
\includegraphics[width=\columnwidth]{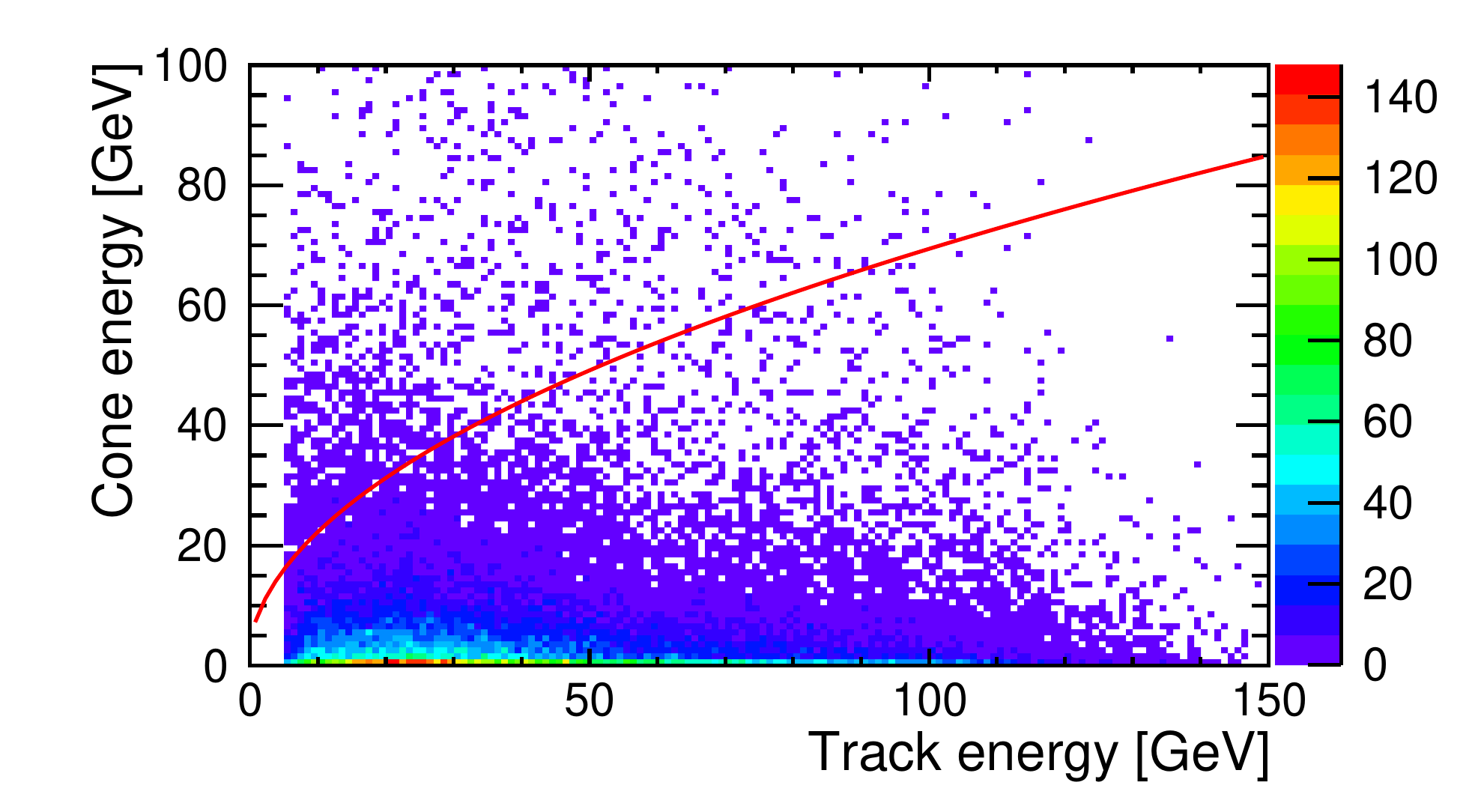}
\begin{textblock}{0.08}(4.5, -2.4)
\textbf{CLICdp}
\end{textblock}
 \end{subfigure}
\quad
\begin{subfigure}{\columnwidth}  
\caption{}\label{fig:4b}
\includegraphics[width=\columnwidth]{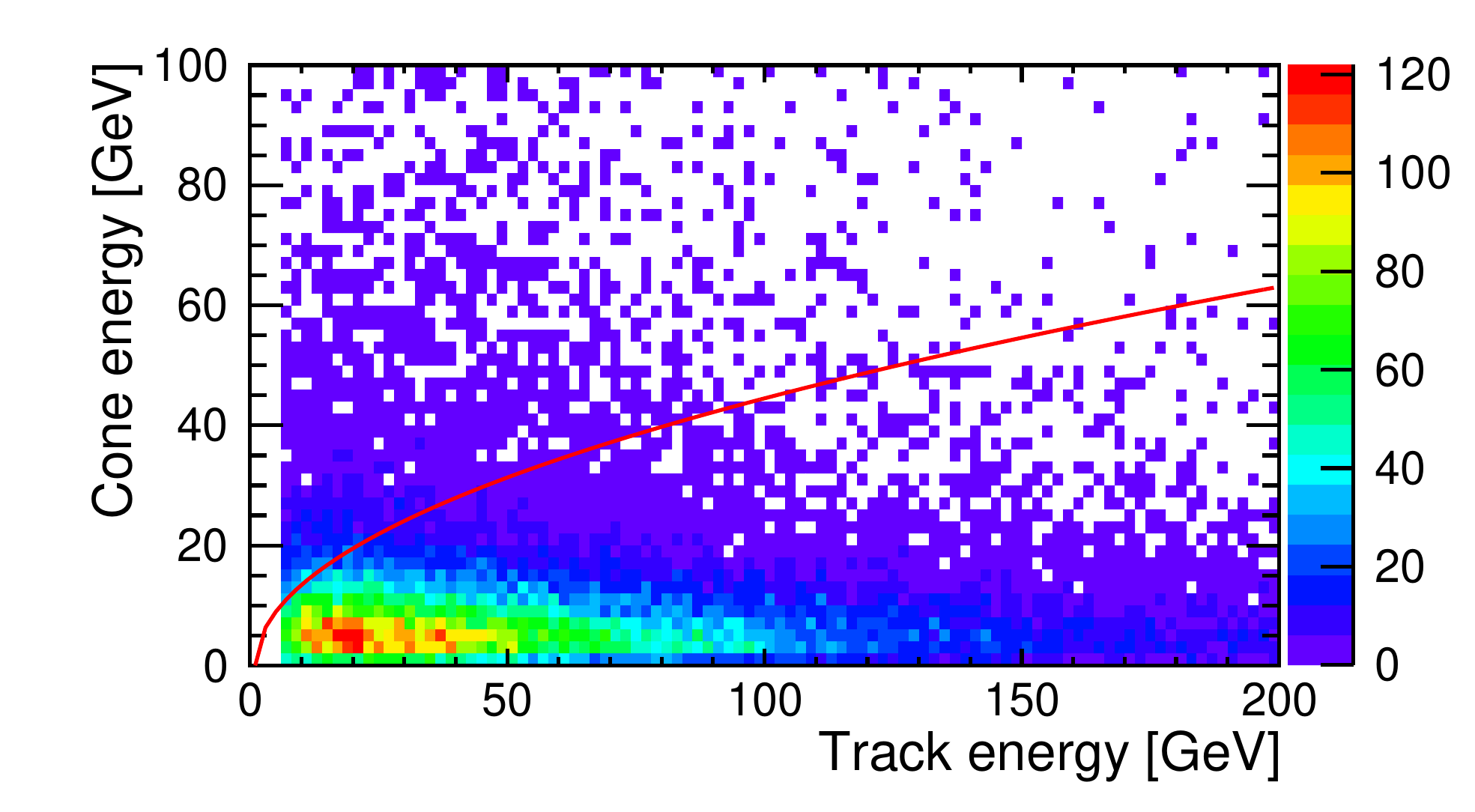}    
\begin{textblock}{0.08}(4.5, -2.4)
\textbf{CLICdp}
\end{textblock}
\end{subfigure}
\caption{\label{fig:4} Cone energy as a function of reconstructed lepton energy at 350\,GeV (a) and at 3\,TeV centre-of-mass energy (b). The red line represents the polynomial distribution from Eq.\ref{eq-Econe}, separating the isolated lepton from other particles in an event. }
\end{figure}

Finally, the leptons from the signal are required to be isolated from other activity within an event. Lepton tracks are required to satisfy two-dimensional requirements on cone energy vs. lepton energy, where the cone energy sums up all particle energies, in a cone size of approximately 6$^{\circ}$ around the isolated lepton track. The isolation requirement is:

\begin{equation}
E_{\mathrm cone}^{\mathrm{2}} < B \cdot E_{trk} + C
\label{eq-Econe}
\end{equation}
\noindent
where $E_{trk}$ and $E_{cone}$ are lepton energy and cone energy, respectively, while the parameters $B$ and $C$ optimized to achieve efficient isolation of signal leptons are found to be: $B$ = 48 GeV and $C$ = 16 GeV$^{2}$ at 350 GeV and $B$ = 20 GeV and $C$ = -20 GeV$^{2}$ at 3 TeV centre-of-mass energies. Figure \ref{fig:4} shows the energy within a cone size of 6$^{\circ}$around a lepton track, as a function of a lepton energy, at 350 GeV (Figure \ref{fig:4a}) and 3 TeV (Figure \ref{fig:4b}).

PFOs that are not identified as isolated leptons are clustered into jets. This is achieved using the FastJet implementation of the $k_{\mathrm{T}}$ algorithm. Events are forced into four (two) jets at 350 GeV (3 TeV) centre-of-mass energy. The distance parameter $R$ corresponding to the effective jet width is chosen to be 1.1 at 350 GeV and 0.7 at 3 TeV. Reconstructed leptons and jets are combined to form $Z$ boson candidates. At 3 TeV, the di-jet or di-lepton with the higher invariant mass is considered to be an on-shell $Z$ boson, while the other fermionic pair forms the off-shell ${Z^\ast}$ boson. At 350\,GeV the invariant mass combination of di-jet and di-lepton pairs that is closest to the simulated Higgs boson mass (126 GeV) is taken as a Higgs candidate with the other pair of quarks considered as the radiated (primary) $Z$ boson\footnote{Though the proposed reconstruction of primary $Z$ and Higgs boson is rather simple, in comparison to the usual $\chi^2$ minimisation of difference of the reconstructed invariant masses w.r.t. the nominal ones \cite{r2}, the method works well since the distribution of difference of the selected reconstructed Higgs boson masses and the generated one is rather narrow (RMS $<$ 5\,GeV), so choice of the combination with closest-to-minimal mass difference seems optimal.}. 

With the criteria described above, the preselection efficiencies for the signal are 77\% and 67\% at 350\,GeV and 3\,TeV, respectively. Preselection efficiences for signal and background processes are given in Table \ref{table:2a} at 350 GeV and in Table \ref{table:2b} at 3 TeV. Signal efficiencies of the isolation curves are 93\% and 86\% at 350 GeV and 3 TeV centre-of-mass energies. This is due to the fact that isolation efficiency is smaller at 3\,TeV than at 350\,GeV, since events at higher centre-of-mass energies are more contaminated with the Beamstrahlung products.

In order to take into account Bremsstrahlung of the final state leptons, energies of photons in a cone of 3$^{\circ}$ around lepton candidate are combined with the charged lepton. This is carried out before any preselection. This correction does not have a significant impact on preselection efficiencies, while it improves the mass resolution of the $Z$ reconstruction and consequently of the MVA performance.  

\begin{figure}
\begin{subfigure}{\columnwidth} 
\caption{}\label{fig:5a}
\label{fig:figpres}
\centering
\includegraphics[width=1.5\linewidth]{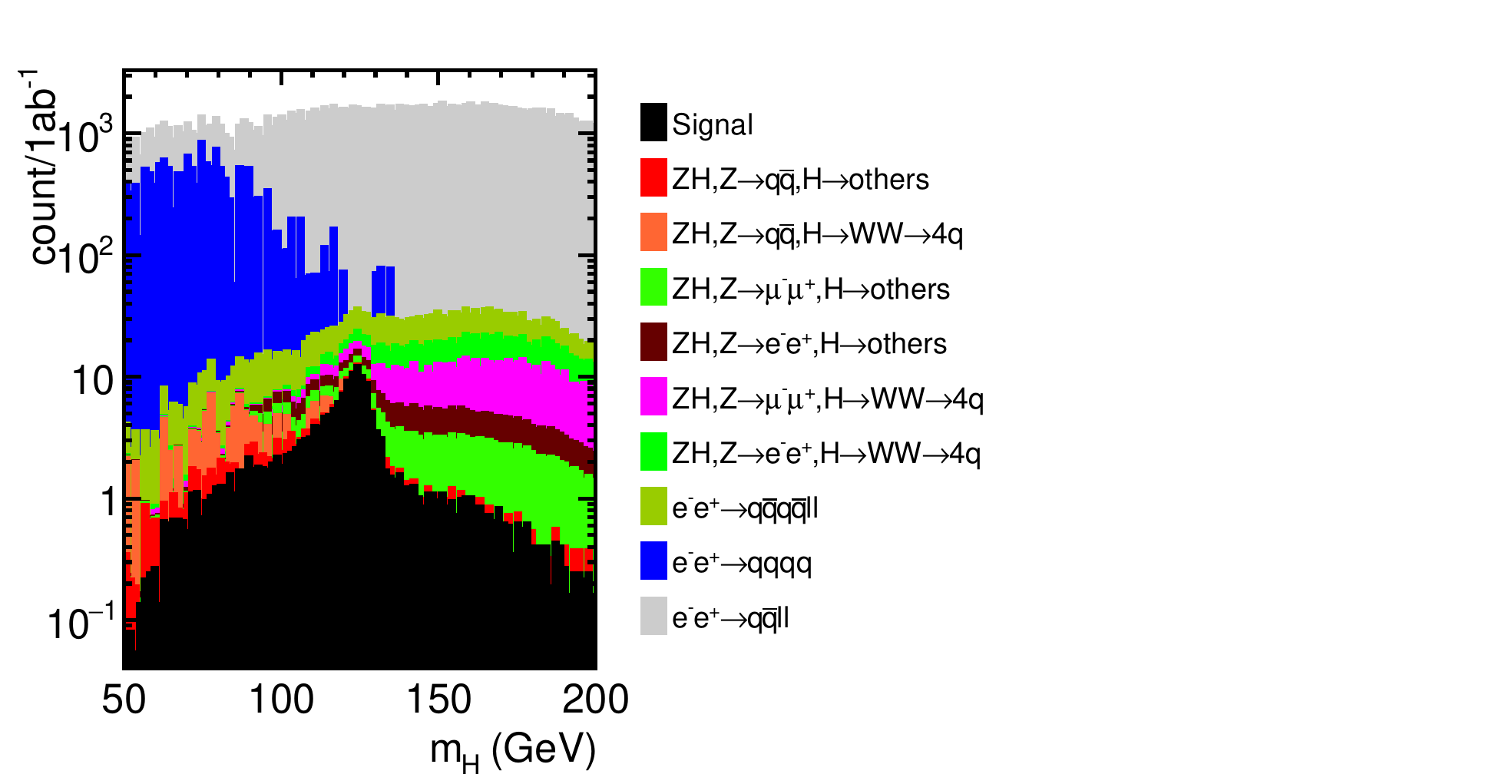}  
\begin{textblock}{0.08}(2.7, -3.4) 
\textbf{CLICdp}
\end{textblock}
\end{subfigure}
\quad
\begin{subfigure}{\columnwidth}
\caption{}\label{fig:5b}
\includegraphics[width=1.5\linewidth]{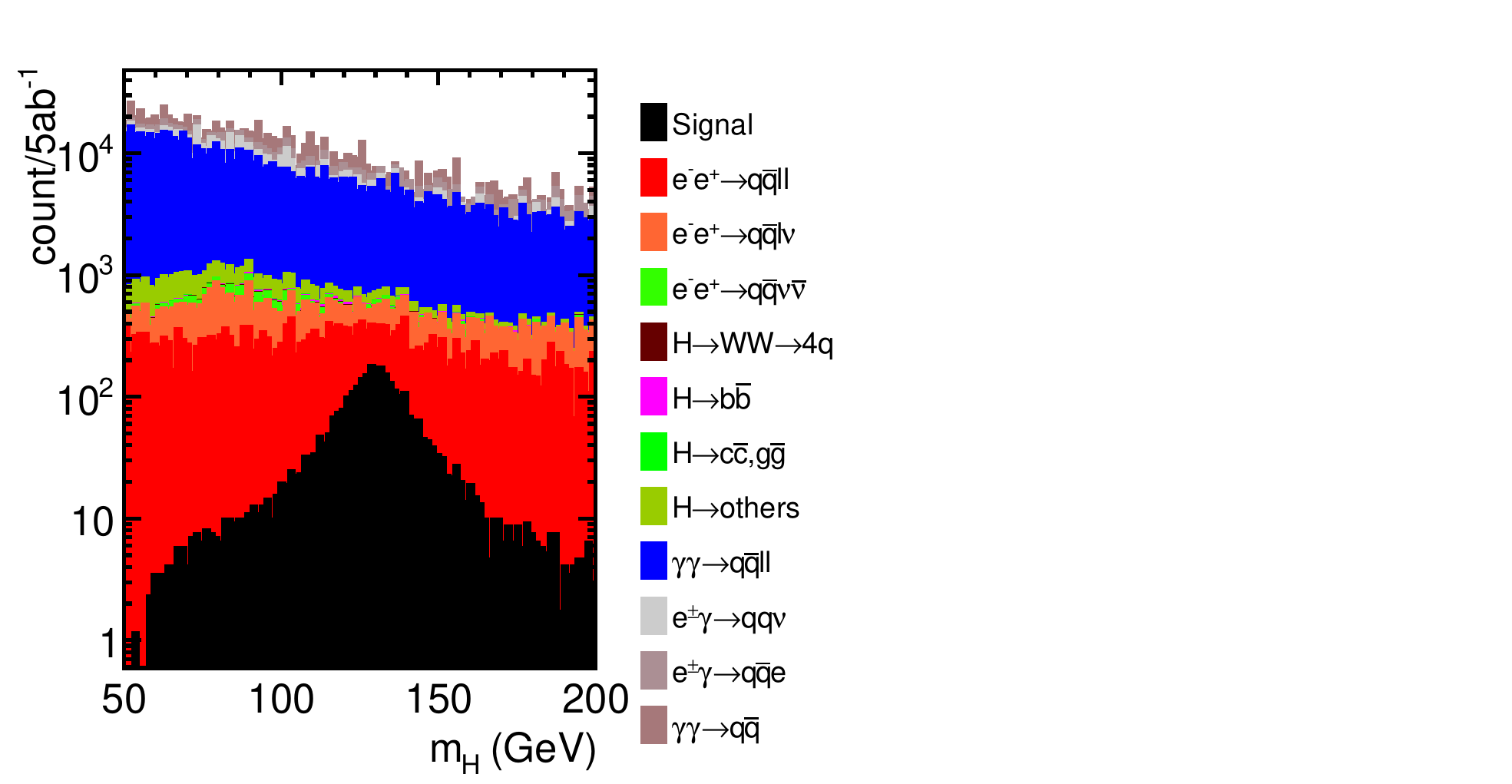}  
\begin{textblock}{0.08}(2.7, -3.4) 
\textbf{CLICdp}
\end{textblock}
\end{subfigure}
\caption{Stacked histograms of the Higgs mass distributions after preselection phase, at 350 GeV (a) and 3 TeV (b).\label{fig:5}}
\end{figure}

In Figure \ref{fig:5}, histograms for signal and background are given for preselected events. Figures \ref{fig:5a} and \ref{fig:5b} show the Higgs mass distributions from the reconstructed $Z$ bosons at 350\,GeV and 3\,TeV, respectively. Background rejection rates are around 97\% and 99.97\% at 350\,GeV and 3\,TeV centre-of-mass energies, respectively. 

\begin{table}[!h]
\centering
\caption{\label{table:2}Summary of preselection efficiencies for signal and background with number of events that pass preselection ($N_{\mathrm{presel}}$), in the considered samples and with expected integrated luminosities at 350 GeV (a) and 3 TeV (b).}
\begin{subtable} {\columnwidth}
\caption{\label{table:2a}}
\begin{tabular*}{\columnwidth}{@{\extracolsep{\fill}}llllll@{}}
\hline
\multicolumn{1}{@{}l}{Signal process}    & {$\epsilon_{\mathrm{presel}} ($\%$) $}  & {$N_{\mathrm{presel}} @ 1 \mathrm{ab^{-1}}$} & {$N_{\mathrm{presel}}$} \\
\hline 
\begin{tabular}{p{3.5cm}p{3cm}}$e^-e^+\rightarrow HZ; Z\rightarrow q \bar{q}, H\rightarrow ZZ^\ast, ZZ^\ast\rightarrow q\bar{q}l^+l^- (l = e, \mu)$\end{tabular}              & 77	& 185	& 13645  \\
\hline
\multicolumn{1}{@{}l}{Background processes}    & {$\epsilon_{\mathrm{presel}} ($\%$) $}  & {$N_{\mathrm{presel}} @ 1 \mathrm{ab^{-1}}$} & {$N_{\mathrm{presel}}$}   \\
\hline 
\begin{tabular}{p{3cm}p{3cm}}$e^-e^+\rightarrow HZ; Z\rightarrow q\bar{q}, H\rightarrow others$\end{tabular}                     &  0.37	& 26	& 285     \\
\begin{tabular}{p{3cm}p{3cm}}$e^-e^+\rightarrow HZ; Z\rightarrow q\bar{q}, H\rightarrow WW\rightarrow 4q$\end{tabular}       & 0.42	& 44 & 50   \\
\begin{tabular}{p{3cm}p{3cm}}$e^-e^+\rightarrow HZ; Z\rightarrow \mu^{+}\mu^{-}, H\rightarrow others$\end{tabular}                      & 61	& 1421	& 51850     \\
\begin{tabular}{p{3cm}p{3cm}}$e^-e^+\rightarrow HZ; Z\rightarrow e^{+}e^{-}, H\rightarrow others$\end{tabular}                      & 62	& 1445	& 52700  \\
\begin{tabular}{p{3cm}p{3cm}}$e^-e^+\rightarrow HZ; Z\rightarrow \mu^{+}\mu^{-}, H\rightarrow WW\rightarrow 4q$\end{tabular}                        & 60	& 434	& 8400  \\
\begin{tabular}{p{3cm}p{3cm}}$e^-e^+\rightarrow HZ; Z\rightarrow e^{+}e^{-}, H\rightarrow WW\rightarrow 4q$\end{tabular}                      & 60	& 434	& 8400   \\
\begin{tabular}{p{3cm}p{3cm}}$e^-e^+\rightarrow q\bar{q}q\bar{q}l^+l^-$\end{tabular}      	& 21	& 939	& 9240    \\
\begin{tabular}{p{3cm}p{3cm}}$e^-e^+\rightarrow qqqq$\end{tabular}             & 0.32	& 18560	&  611    \\
\begin{tabular}{p{3cm}p{3cm}}$e^-e^+\rightarrow q\bar{q}l^+l^-$\end{tabular}              & 11.4	& 193800	& 85044   \\
\hline
\hline
\end{tabular*}
\end{subtable}
\quad
\begin{subtable} {\columnwidth}
\vspace*{0.5 cm}
\caption{\label{table:2b}}
\begin{tabular*}{\columnwidth}{@{\extracolsep{\fill}}llllll@{}}
\hline
\multicolumn{1}{@{}l}{Signal processes}    & {$\epsilon_{\mathrm{presel}} (\%)$}	& {$N_{\mathrm{presel}} @ 5 \mathrm{ab^{-1}}$} & {$N_{\mathrm{presel}}$}    \\
\hline 
\begin{tabular}{p{3cm}p{3cm}}$e^-e^+\rightarrow H\nu\bar{\nu}; H\rightarrow ZZ^\ast, ZZ^\ast\rightarrow q\bar{q}l^+l^-,\break (l = e, \mu)$\end{tabular}              & 67	& 3788	& 11224     \\
\hline
\multicolumn{1}{@{}l}{Background process}    & {$\epsilon_{\mathrm{presel}} (\tcperthousand) $}	& {$N_{\mathrm{presel}} @ 5 \mathrm{ab^{-1}}$} & {$N_{\mathrm{presel}}$}   \\
\hline 
\begin{tabular}{p{3cm}p{3cm}}$e^-e^+\rightarrow H\nu\bar{\nu}; H\rightarrow WW, WW\rightarrow 4q$\end{tabular}       & 1.7	& 371	& 372  \\
\begin{tabular}{p{3cm}p{3cm}}$e^-e^+\rightarrow H\nu\bar{\nu}; H\rightarrow b\bar{b}$\end{tabular} 	& 0.6	& 720	& 660    \\
\begin{tabular}{p{3cm}p{3cm}}$e^-e^+\rightarrow H\nu\bar{\nu}; H\rightarrow c\bar{c}$\end{tabular}  & 0.6	& 35	& 31  \\
\begin{tabular}{p{3cm}p{3cm}}$e^-e^+\rightarrow H\nu\bar{\nu}; H\rightarrow gg$\end{tabular}      & 0.9		& 158	& 115  \\
\begin{tabular}{p{3cm}p{3cm}}$e^-e^+\rightarrow H\nu\bar{\nu}; H\rightarrow others$\end{tabular}  & 45	& 20340	& 20925  \\
\begin{tabular}{p{3cm}p{3cm}}$e^-e^+\rightarrow q\bar{q}l^+l^-$\end{tabular}                      & 7.5	& 124500	& 15000   \\
\begin{tabular}{p{3cm}p{3cm}}$e^-e^+\rightarrow qql\nu$\end{tabular}            & 3	& 83400	& 9300    \\
\begin{tabular}{p{3cm}p{3cm}}$e^-e^+\rightarrow q\bar{q}\nu\bar{\nu}$\end{tabular}  &  0.7	&  4620	& 398   \\
\begin{tabular}{p{3cm}p{3cm}}$\gamma\gamma\rightarrow q\bar{q}l^+l^-$\end{tabular}       & 11	& 1500000	& 27500    \\
\begin{tabular}{p{3cm}p{3cm}}$\gamma\gamma\rightarrow q\bar{q}$\end{tabular}  & 1	& 517400	& 1000   \\
\begin{tabular}{p{3cm}p{3cm}}$e^\pm\gamma\rightarrow q\bar{q}e$\end{tabular}   & 8.8	& 530640	& 4066     \\
\begin{tabular}{p{3cm}p{3cm}}$e^\pm\gamma\rightarrow qq\nu$\end{tabular}    & 1.4	& 193620	&  968   \\
\hline
\hline
\end{tabular*}
\end{subtable}
\end{table} 

\section{\label{sec:leve5}Multivariate analysis}
\subsection{\label{subsec:leve5.1} MVA at 350 GeV }
Separation of signal from background uses a multivariate analysis based on the Boosted Decision Trees (BDT) classifier \cite{r21}. At 350 GeV, an MVA is trained with the following observables: mass of the on-shell $Z$ boson; mass of the off-shell $Z$ boson; mass of the primary $Z$; invariant mass of two selected leptons; invariant mass of two reconstructed jets; mass of a Higgs candidate; visible energy in the event; difference between the visible energy and the Higgs energy; polar angle of a Higgs candidate; angle between on-shell and off-shell $Z$ bosons in the plane perpendicular to the beam axis; number of all PFO objects in an event; jet transition variables ($-\log y_{12}$,$-\log y_{23}$ and $-\log y_{45}$); b-tag and c-tag probabilities of jets sorted by decreasing transverse momentum of a jet; transverse momenta and energies of isolated leptons. Individual leptons are sorted in a way such that the higher transverse momentum lepton has index 1. The Higgs mass is constrained in the interval (50 GeV $< m_{H} < $ 170 GeV). At both centre-of-mass energies Higgs mass window is chosen to selects intervals where signal is naturally present with reasonable statistics. The three most sensitive observables in the BDT training phase are found to be: energy of the reconstructed lepton with the highest $p_{\mathrm{T}}$, jet transition variable ($-\log y_{23}$) and mass of the reconstructed primary $Z$.

The BDT output variable cut-off value is chosen to  maximize the statistical significance $S$:

\begin{equation}
\label{sig}
S = N_{S}/ \sqrt{N_{S}+N_{B}} 
\end{equation}
where $N_{S, B}$ denotes the number of selected signal and background events. Relative statistical uncertainty $\delta$ is derived from the statistical significance as $\delta = 1/S$. The optimal BDT cut is found to be 0.20, corresponding to a statistical significance of 5. The overall efficiency of the signal including preselection and MVA selection is found to be approximately 19\%, due to the relatively low MVA efficiency of approximately 25\%.
The uncertainty of the estimated number of signal and background events in 1 ab$^{-1}$ of data leads to the 2\% uncertainty of our estimate of $\delta$ ($\delta$ = (20 $\pm$ 2)\%), from the Poisson variance of the number of selected background and signal events. Histograms of the Higgs mass distributions for signal and background after all selection phases are given in Figure \ref{fig:6a}.

\subsection{ \label{subsec:leve5.2} MVA at 3 TeV }
At 3 TeV centre-of-mass energy, the MVA is trained with the following observables: mass of the on-shell $Z$ boson; mass of the off-shell $Z$ boson; invariant mass of two selected leptons; invariant mass of two reconstructed jets; mass of a Higgs candidate; visible energy in an event; difference between the visible energy and the Higgs energy; polar angle of a Higgs candidate; missing transverse momentum per event; number of all PFO objects in an event; jet transition variables ($-\log y_{12}$ and $-\log y_{23}$); b-tag and c-tag probabilities of jets sorted by decreasing transverse momentum of a jet. The Higgs candidate mass is limited to the interval (75\,GeV $< m_{H} <$ 175\,GeV). The three most sensitive observables are found to be masses of Higgs and off-shell bosons and polar angle of the reconstructed Higgs boson.  
 
\begin{figure}[!h]
\begin{subfigure}{\columnwidth}   
\label{fig:fig1a2} 
\caption{}\label{fig:6a} 
\centering
\includegraphics[width=1.5\linewidth]{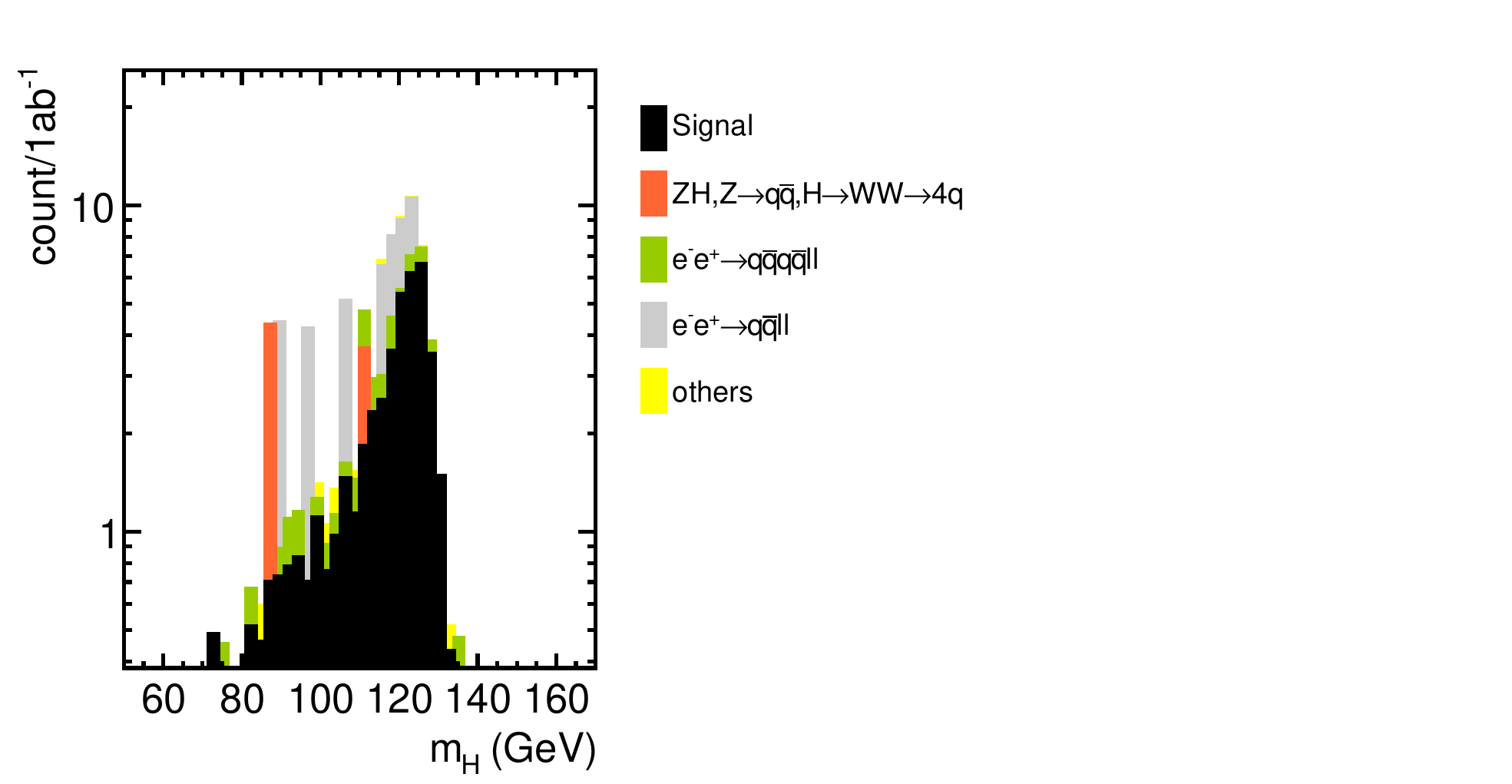} 
\begin{textblock}{0.08}(2.7, -3.4)
\textbf{CLICdp}
\end{textblock} 
\end{subfigure}
\quad
\begin{subfigure}{\columnwidth} 
\caption{}\label{fig:6b}
\includegraphics[width=1.5\linewidth]{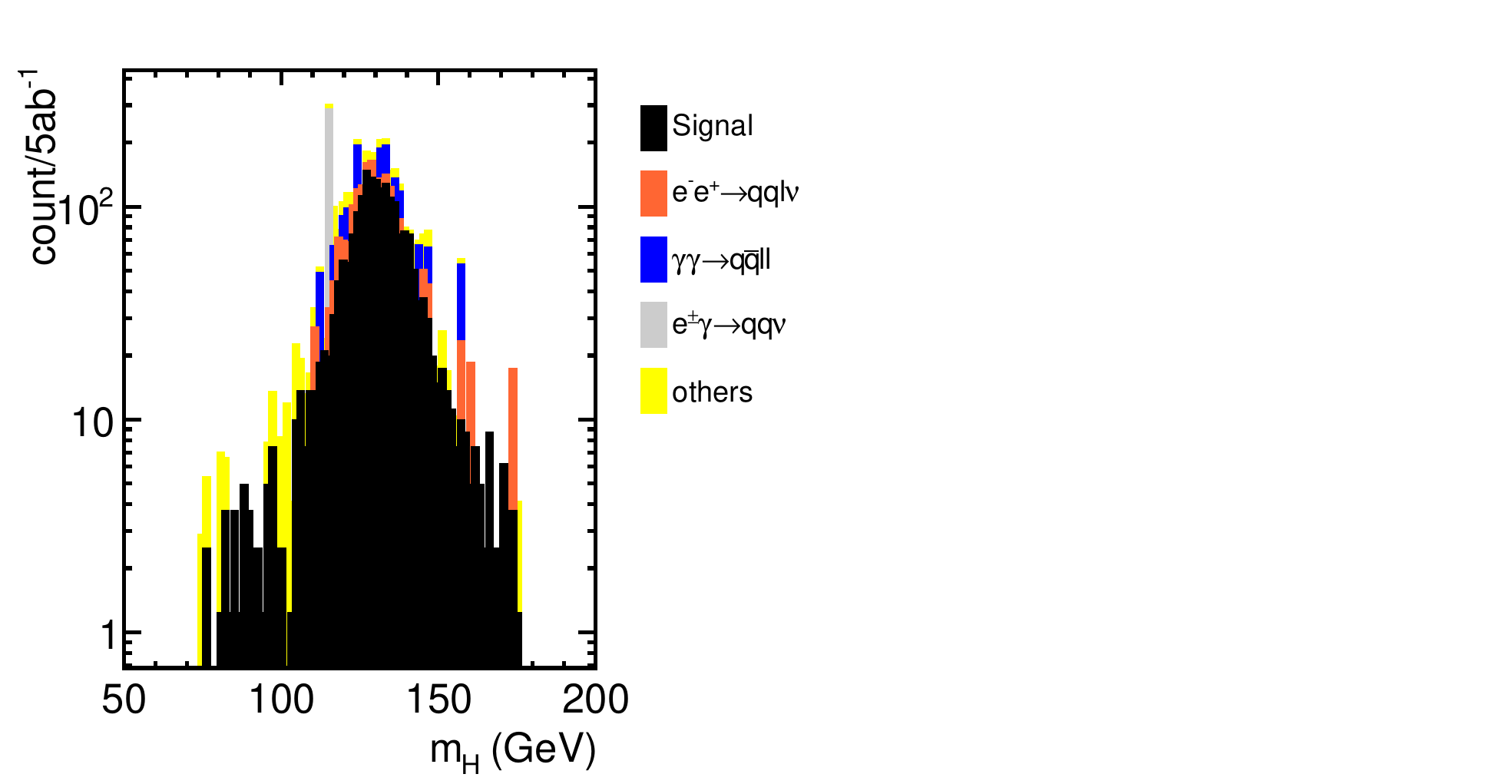}    
\begin{textblock}{0.08}(2.7, -3.4)
\textbf{CLICdp}
\end{textblock}
\end{subfigure}
\caption{Stacked histograms of the Higgs mass distributions after MVA, at 350 GeV (a) and 3 TeV (b) centre-of-mass energies\label{fig:6}.}
\end{figure}

The optimal BDT cut is found to be 0.11, corresponding to a statistical significance of 33. The overall efficiency of signal selection including preselection and MVA selection is found to be about 36\%. This corresponds to the MVA signal selection efficiency of approximately 53\%. Figure \ref{fig:6b} presents the Higgs mass distributions for signal and background after MVA selection. The BDT background efficiency is on average at the permille level and Table \ref{table:3} gives the composition of irreducible backgrounds. 

\vspace{10mm}

\begin{table}[!h]
\centering
\caption{\label{table:3} Preselection and MVA selection efficiences for signal and irreducible background processes and number of selected events $N_{\mathrm{BDT}}$, at 3 TeV centre-of-mass energy, in 5 ab$^{-1}$ of data.}
\begin{tabular*}{\columnwidth}{@{\extracolsep{\fill}}llllll@{}}
\hline
\begin{tabular}{p{4cm}p{8cm}} Process \end{tabular}   & {$\epsilon_{\mathrm{presel}}$}	& {$\epsilon_{\mathrm{BDT}}$} & {$N_{\mathrm{BDT}}$}     \\
\hline
\begin{tabular}{p{3cm}p{2cm}} Signal $@$ 3 TeV \end{tabular} 	& 67\%	& 53\%	& 2020    \\ 
\hline
\begin{tabular}{p{4cm}p{8cm}} Background processes  \end{tabular}    \\
\hline
\begin{tabular}{p{3cm}p{3cm}}$\gamma\gamma\rightarrow q\bar{q}l^+l^-$\end{tabular}       & 11$\tcperthousand$	& 0.3$\tcperthousand$	& 438    \\
\begin{tabular}{p{3cm}p{3cm}}$e^-e^+\rightarrow qql\nu$\end{tabular}            & 3$\tcperthousand$	& 4$\tcperthousand$  & 322    \\
\begin{tabular}{lp{3cm}p{3cm}}$e^-e^+\rightarrow H\nu\bar{\nu}; H\rightarrow others$\end{tabular}  & 45$\tcperthousand$	& 1.3\%	& 259  \\
\begin{tabular}{p{3cm}p{3cm}}$e^\pm\gamma\rightarrow qq\nu$\end{tabular}    & 8.8$\tcperthousand$	& 1.3$\tcperthousand$	&  252	\\ 
\begin{tabular}{lp{3cm}p{3cm}} processes with $N_{BDT} <$ 100  \end{tabular}    & 5.3$\tcperthousand$	& 1.1$\tcperthousand$	&  140   \\
\hline
\hline
\end{tabular*}
\end{table}

\section{\label{sec:leve6} Statistical uncertainties}
As said in Section~\ref{subsec:leve5.1}, the relative statistical uncertainty of the ${\sigma(H\nu\bar{\nu})\times BR(H\rightarrow ZZ^\ast)}$ measurement is derived from statistical significance.
The uncertainty of the estimated number of signal and background events at 350 GeV, with 1 ab$^{-1}$ of data, leads to the 2\% uncertainty of our estimate of $\delta$ ($\delta$ = (20 $\pm$ 2)\%). The uncertainty of the number of background events at 3\,TeV is obtained in the same way as discussed in Section~\ref{subsec:leve5.1}. With 5 ab$^{-1}$ of data, uncertainty of our estimate of $\delta$ is 0.1\% ($\delta$ = (3.0 $\pm$ 0.1)\%). 
The high-energy result can be further improved by the beam polarization due to the chiral nature of $WW-$fusion. Assuming the beam polarization scheme discussed in Section~\ref{sec:leve3}, the statistical uncertainty of the 3 TeV measurement can be decreased by a factor of $\sim \sqrt{1.5}$ \cite{r8}.

\section{\label{sec:leve7}Conclusions}
The statistical precision of the measurement of ${\sigma(H\nu\bar{\nu})\times BR(H\rightarrow ZZ^\ast)}$ at CLIC, using data from 350\,GeV and 3\,TeV centre-of-mass energies is determined on the basis of a full simulation of physics processes and detector response. Both measurements are carried out using the semi-leptonic signal final states. The relative statistical uncertainty of ${\sigma(H\nu\bar{\nu})\times BR(H\rightarrow ZZ^\ast)}$ is found to be 20\% at 350\,GeV and 3.0\% at 3\,TeV, assuming integrated luminosities of 1\,ab$^{-1}$ and 5\,ab$^{-1}$, respectively. The statistical uncertainty at 3 TeV is consistent with the expectations from \cite{r2} based on luminosity scaling of the precision of a 1.4\,TeV measurement. The statistical uncertainty of the high-energy result can be further reduced through enhancement of the signal with the proposed beam polarization scheme. 

However, the ultimate sub-percent precision of the Higgs to Z bosons coupling will be obtained from a global fit of individual measurements as the ones discussed in this paper, combined in a model-independent or model-dependent way \cite{r2}.

\begin{acknowledgments}
The work presented in this paper has  been carried out in the framework of the CLIC detector and physics study (CLICdp) collaboration and the authors would like to thank CLICdp members for their support, in particular to the colleagues from the Analysis Working Group for useful discussions. We are particularly grateful to Aleksander Filip \.{Z}arnecki and Philipp Roloff, for useful ideas exchanged in the course of the analysis and to Nigel Watson for improving the text in various aspects. We acknowledge the support received until 2020, from the Ministry of Education, Science and Technological Development of the Republic of Serbia within the national project OI171012. 
\end{acknowledgments}


\begin{thebibliography}{99}
\bibitem{r1}
R. S. Gupta, H. Rzehak, J. D. Wells, How well do we need to measure Higgs boson couplings, Phys. Rev. D 86 \href{http://link.aps.org/doi/10.1103/PhysRevD.86.095001} {095001} (2012), \href{http://arxiv.org/abs/1206.3560} {arXiv:1206.3560}

\bibitem{r2}
H. Abramowicz et al. [CLICdp Collaboration], Higgs physics at the CLIC Electron-Positron Linear Collider, Eur. Phys. J. C 77, 475 (2017), \href{https://arxiv.org/abs/1608.07538}{arXiv:1608.07538}

\bibitem{goca}
G. Milutinovi\'{c}-Dumbelovi\'{c}, Methods of the ${\sigma\times BR(H\rightarrow \mu^{+}\mu^{-})}$ and ${\sigma\times BR(H\rightarrow ZZ^\ast)}$ measurements at 1.4 TeV CLIC, PhD thesis, University of Belgrade (2017), \href{http://cds.cern.ch/record/2312030}{CERN-THESIS-2017-349}

\bibitem{r3}
J. Ellis, P. Roloff, V. Sanz and T. You, Dimension-6 Operator Analysis of the CLIC Sensitivity to New Physics, KCL-PH-TH/2017-04, CERN-PH-TH/2017-009, Cavendish-HEP-17/01, DAMTP-2017-01 (2017), \href{https://arxiv.org/abs/1701.04804} {arXiv:1701.04804}

\bibitem{r4}
L. Linssen, A. Miyamoto, M. Stanitzki, H. Weerts (eds.), Physics and Detectors at CLIC: CLIC Conceptual Design Report, ANL-HEP-TR-12-01, CERN-2012-003, DESY 12-008, KEK Report 2011-7 (2012), \href{https://arxiv.org/abs/1202.5940}{ arXiv:1202.5940 }

\bibitem{r5}
T. Abe et al., The International Large Detector: Letter of Intent, DESY-2009-87, FERMILAB-PUB-09-682-E, KEK-REPORT-2009-6 (2010), \href{http://arxiv.org/abs/1006.3396}{arXiv:1006.3396}

\bibitem{r6}
N. Alipour Tehrani et al., CLICdet: The post-CDR CLIC detector model, CLICdp-Note-2017-001 (2017),  \href{https://cds.cern.ch/record/2254048/files/CLICdp-Note-2017-001.pdf}{https://cds.cern.ch/record/2254048}

\bibitem{r7}
M. A. Thomson,
Particle Flow Calorimetry and the Pandora PFA Algorithm, Nucl. Instrum. Methods A 611, 25 (2009), \href{https://arxiv.org/abs/0907.3577}{ arXiv:0907.3577 }


\bibitem{r8}
A. Robson and P. Roloff, Updated CLIC luminosity staging baseline and Higgs coupling prospects, CLICdp-Note-2018-002 (2018), \href{https://arxiv.org/abs/1812.01644}{arXiv:1812.01644}

\bibitem{r9}P. Roloff, R. Franceschini, U. Schnoor, A.Wulzer (eds.), The Compact Linear $e^{+}e^{-}$ Collider (CLIC): Physics Potential, Input to the European Particle Physics Strategy Update on behalf of the CLIC and CLICdp Collaborations (2018), \href{https://arxiv.org/abs/1812.07986}{arXiv:1812.07986}


\bibitem{r10}S. Dittmaier et al., Handbook of LHC Higgs Cross Sections: 2. Differential Distributions, CERN-2012-002 (2012), \href{https://arxiv.org/abs/1201.3084}{arXiv:1201.3084} 


\bibitem{r11}
W. Kilian, T. Ohl, J. Reuter, WHIZARD: Simulating Multi-Particle Processes at LHC and ILC, Eur. Phys. J. C 71, 1742 (2011), \href{http://arxiv.org/abs/0708.4233} {arXiv:0708.4233}

\bibitem{r12}
T. Sjostrand, S. Mrenna, P. Z. Skands,
PYTHIA 6.4 Physics and Manual, JHEP 05, 026 (2006), \href{http://arxiv.org/abs/hep-ph/0603175}{arXiv:hep-ph/0603175}

\bibitem{r13}
D. Schulte, Beam-beam simulations with GUINEA-PIG, (1999), \href{http://cds.cern.ch/record/382453}{CERN-PS-99-014-LP}

\bibitem{r14}
P. Mora de Freitas, H. Videau, Detector Simulation with Mokka/Geant4: Present and Future, International Workshop on Linear Colliders, JeJu Island, Korea (2002), \href{http://inspirehep.net/record/609687?ln=en}{LC-TOOL-2003-010}

\bibitem{r15}
S. Agostinelli et al., Geant4 - A Simulation Toolkit, Nucl. Instrum. Methods Phys. Res., Sect. A 506, 3 (2003)

\bibitem{r16}
J. Marshall, A. M\"{u}nnich, M. Thomson, Performance of Particle Flow Calorimetry at CLIC, Nucl. Instrum. Methods A 700, 153 (2013), \href{http://arxiv.org/abs/1209.4039}{arXiv:1209.4039}

\bibitem{r17}
S. Catani et al., Longitudinally-invariant $k_{\bot}-$clustering algorithms for hadron-hadron collisions, Nucl. Phys. B 406, 187 (1993)

\bibitem{r18}
G. S. M. Cacciari, G. Soyez, FastJet User Manual, Eur. Phys. J. C 72, 1896 (2012),
\href{https://arxiv.org/pdf/1111.6097v1.pdf}{arXiv:1111.6097}

\bibitem{r19}
O. Wendt, F. Gaede, T. Kramer, Event reconstruction with MarlinReco at the ILC, Pramana 69 (2007),  \href{http://arxiv.org/abs/physics/0702171}{arXiv:physics/0702171}

\bibitem{r20}
T. Suehara, T. Tanabe, LCFIPlus: A Framework for Jet Analysis in Linear Collider Studies, Nucl. Instrum. Meth. A 808, 109-116 (2016)

\bibitem{r21}
A. H\"{o}cker et al., TMVA - Toolkit for multivariate data analysis (2009), \href{http://arxiv.org/abs/physics/0703039}{arXiv:physics/0703039}


\bibitem{r22}
C. Grefe et al., ILCDIRAC, a DIRAC extension for the Linear Collider community,
CLICdp-Conf 2013-003, CERN, Geneva (2013), \href{http://cds.cern.ch/record/1626585/files/ChepProceedings.pdf}{https://cds.cern.ch/record/1626585/}


 
\end{thebibliography}
\end{document}